\documentclass[a4paper,12pt]{article}
\usepackage{amsmath,amsfonts,amssymb,cite}
\begin{document}

\begin{center}
{\Large\bf Contribution of higher meson resonances to
the electromagnetic $\pi$-meson mass difference}\\

\vspace{4mm}

{\bf V. A. Andrianov, S. S. Afonin\footnotemark[1]}\footnotetext[1]{E-mail:
Sergey.Afonin@pobox.spbu.ru}
\\
\smallskip
V.A. Fock Department of Theoretical Physics, St.Petersburg State
University, 198504, St.Petersburg, Russia
\end{center}

\begin{abstract}

Modifications of the DGMLY relation for
calculation of electromagnetic $\pi$-meson mass difference based on the Chiral Symmetry
Restoration phenomenon at high energies as well as
the Operator Product Expansion of quark densities for vector ($\rho$) and
axial-vector ($a_1$) meson fields difference are proposed.
In the calculations higher meson resonances in vector and axial-vector channels
are taken into account. It is shown that the inclusion of the first $\rho$
and $a_1$ radial excitations improves the results for electromagnetic
$\pi$-meson mass
difference as compared with the previous ones. Estimations on the
electromagnetic $\rho$ and $a_1$-meson decay constants and the constant $L_{10}$ of
effective chiral Lagrangian are obtained from the generalized Weinberg sum rules.

\end{abstract}

\section{Introduction}

It is well known that the correlation functions of vector (V) and axial-vector (A) meson fields
are connected with some experimentally observed characteristics
of pseudoscalar mesons, in particular, with
the electromagnetic $\pi$-meson mass difference $\Delta m_{\pi}|_{em}$~\cite{37}.
Recent experimental data of ALEPH~\cite{57} and OPAL~\cite{57d} collaborations
on hadronic $\tau$-decays ($\tau\rightarrow(VA)\nu_{\tau}$,
$\tau\rightarrow\pi\nu_{\tau}$) point out that one should  take into account
more degrees of freedom in pseudoscalar channels
for the (VV)-(AA) - correlators in order to check both perturbative and
non-perturbative QCD parameters. In particular, the authors of~\cite{53}
fulfilled an analysis of ALEPH~\cite{57} experimental data to
measure the correlator difference
$\Pi^V-\Pi^A$ behaviour from the decay $\tau\rightarrow (V,A)\nu_{\tau}$.
This correlator difference happens to vanish practically already at
intermediate energies $\lesssim3$~GeV.
A small contribution of the first radial excitations of vector mesons to the $\Pi^V-\Pi^A$
can be seen experimentally, but a contribution of following ones is nearly negligible.

As it follows from~\cite{37, 54} the bulk of pion mass difference effect
has, in general, an electromagnetic origin. There are different approaches to calculate
an electromagnetic contribution to
$\Delta m_{\pi}|_{em}$~\cite{54, min, 42, 36, 3a, 41, 38, 43, 44, 33, mous}.
One of simple ways to calculate this quantity in the chiral limit and to the
lowest order in the electromagnetic interactions was proposed in~\cite{37},
where it was calculated in the framework of Current Algebra with the help of
Weinberg sum rules~\cite{Wein1}, the latter being saturated by one vector ($\rho$(770))
and one axial-vector ($a_1$(1260)) mesons. However, from PDG~\cite{31} (see also~\cite{ani}) it is known
that there is a series of heavier meson states with the same quantum numbers
--- $\rho(1450)$, $\rho(2150)$, which represent the radial excitations
of $\rho(770)$-meson in a language of potential quark models. In the axial-vector channel
one might have radial excitations of $a_1(1260)$-resonance as well. However, their mass spectrum
is not yet accurately identified~\cite{a11, a12, a13}.
%В частности, косвенным
%указанием на их существование является результат линейной
%реализации киральной симметрии, в соответствии с которой, каждый
%$\rho$-мезон или его родственник должен иметь своего кирального
%партнера по четности.
Thus, the problem arises: what is a contribution of these higher meson
resonances to the electromagnetic $\pi$-meson mass difference $\Delta m_{\pi}|_{em}$?

In this work we investigate the contribution of the first radial
(V,A) excitations to the electromagnetic $\pi$-meson mass
difference. The analysis is based on modifications of the
relation~\cite{37} with an additional Weinberg sum
rule, following from the requirement of Chiral Symmetry
Restoration at high energies~\cite{25} and the Operator Product
Expansion (OPE)~\cite{18}, and without one. It should be stressed
that our goal is not to calculate $\Delta m_{\pi}|_{em}$ strictly
(this quantity is known from experiment) but we are going to employ this
difference to calculate some physical parameters of hadron
resonances as well as to investigate a contribution of higher meson resonances
in saturation of physical observables.

The paper is organized as follows. In the section 2 we remind the idea of derivation of
the DGMLY relation for electromagnetic $\pi$-meson mass difference and present a scheme of obtaining
the Weinberg sum rules from the requirement of chiral symmetry restoration at high energies
and OPE. The section 3 is devoted to an extension of the classical formula~\cite{37}
by taking into account of higher meson (V,A) resonances. Here we calculate also certain
electromagnetic decay constants of $\rho$ and $a_1$ mesons and the constant $L_{10}$
of effective chiral Lagrangian~\cite{32}, which can be determined, in particular, from
the decay $\pi\rightarrow e\nu\gamma$. The obtained results are discussed in the section 4.

\section{The electromagnetic $\pi$-meson mass difference and the Weinberg sum rules}

The electromagnetic $\pi$-meson mass difference $\Delta
m_{\pi}|_{em}$ is defined in the chiral limit
$m_{cur}\!\!=0$ by the known relation~\cite{37, 36, 38, 39a}:
\begin{equation}
\label{main}
(m_{\pi^{+}}^{2}-m_{\pi^{0}}^{2})|_{em}=\frac{2e^{2}C}{f_{\pi}^{2}}\,,
\end{equation}
where in~(\ref{main}) the constant $C$ is:
\begin{equation}
\label{c}
C=-\frac{1}{8\pi^{2}}\cdot\frac{3}{4}\int\limits_{0}^{\infty}\!dp^{2}\cdot
p^{2} [\Pi^{V}(p^{2})-\Pi^{A}(p^{2})]\,,
\end{equation}
and
\begin{equation}
\Pi^{C}(p^{2})=\int d^{4}x\exp(ipx)\langle T(\bar{q}(x)\Gamma
q(x)\bar{q}(0)\Gamma q(0)) \rangle\,,
\end{equation}
$$
C\equiv V,A\,; \qquad \Gamma\equiv
\gamma_{\mu},\gamma_{\mu}\gamma_{5}
$$
represent the two-point correlators of vector and axial-vector quark densities
in Euclidean space.
In the large-$N_c$ limit (planar limit) the correlators of colour singlet quark
densities are saturated by narrow meson resonances only~\cite{16,17}:
\begin{equation}
\label{kor}
\Pi^{C}(p^{2})|_{planar}=\sum_{n}\frac{Z_{n}^{C}}{p^{2}+m_{C,n}^{2}}\,.
\end{equation}

On the other hand, owing to the asymptotic freedom of QCD, a high energy asymptotic
of these correlators is described by the perturbation theory and Operator
Product Expansion~\cite{18}. To the lowest order its behaviour is given by:
\begin{equation}
\label{as}
\Pi^C(p^2)|_{p^2\rightarrow\infty}\sim
p^2\ln\!\frac{p^2}{\mu^2}\,,
\end{equation}
where $\mu$ is a normalization point of fermion currents.

From the comparison of~(\ref{kor}) and~(\ref{as}) one can infer that an infinite set of resonances
with equal quantum numbers must exist in order to reproduce the perturbative
asymptotic~(\ref{as}). In the chiral limit and in the large-$N_c$
approach it can be shown that in VA-channels~\cite{25, 20}:
\begin{equation}
\label{A-V}
\left(\Pi^{V}(p^{2})-\Pi^{A}(p^{2})\right)_{p^{2}\rightarrow\infty}\equiv
\frac{\Delta_{VA}}{p^{6}}+\mathcal{O}\!\left(\frac{1}{p^{8}}\right)\,,
\quad \Delta_{VA}\simeq -16\pi\alpha_{s}<\!\bar{q}q\!>^{2},
\end{equation}
where
\begin{equation}
\label{proect}
\Pi_{\mu\nu}^{V,A}(p^{2})\equiv(-\delta_{\mu\nu}p^{2}+p_{\mu}p_{\nu})
\Pi^{V,A}(p^{2})\,.
\end{equation}

As it follows from the relation~(\ref{A-V}), due to a rapid convergence of the difference
$\Pi^{A}-\Pi^{V}$ with increasing~$p^{2}$ one may expect the chiral symmetry restoration
at high energies, with the difference~$\Pi^{A}-\Pi^{V}$ being approximated by an order
parameter of chiral symmetry breaking (CSB) in QCD i.e. the quark condensate $<\!\bar{q}q\!>$.
The asymptotics~(\ref{A-V}) represents the so called condition of chiral symmetry restoration
in VA-channels. It leads to a certain set of sum rules (see, e.g.~\cite{25}) for spectral
characteristics of VA-mesons. Namely, expanding~(\ref{A-V}) in powers of $p^2$ one
arrives at:
\begin{equation}
\label{W1}
\sum_{n}Z_{n}^{V}-\sum_{n}Z_{n}^{A}=4f_{\pi}^{2}\,,
\end{equation}
\begin{equation}
\label{W2}
\sum_{n}Z_{n}^{V}m_{V,n}^{2}-\sum_{n}Z_{n}^{A}m_{A,n}^{2}=0\,,
\end{equation}
\begin{equation}
\label{W3}
\sum_{n}Z_{n}^{V}m_{V,n}^{4}-\sum_{n}Z_{n}^{A}m_{A,n}^{4}=\Delta_{VA}\,.
\end{equation}
Relations~(\ref{W1}) and~(\ref{W2}) are the Weinberg sum rules~\cite{Wein1} where
$f_{\pi}=93$ MeV is a weak~$\pi$-meson decay constant. The equation~(\ref{W3})
represents an additional sum rule following from OPE. The vector and axial-vector
residues are connected to the electromagnetic meson widths through the relation~\cite{19}:
\begin{equation}
\label{resid}
Z_{n}^{(V,A)}=4f_{(V,A),n}^{2}m_{(V,A),n}^{2}\,,
\end{equation}
where $f_{(V,A),n}$ are (dimensionless) electromagnetic decay constants.

The Weinberg sum rules ensure convergence of the integral~(\ref{c})
in the ultraviolet limit. As a result, one obtains:
%\footnote{В случае ненулевой токовой
%массы кварка правая часть второго правила сумм
%Вайнберга~(\ref{W2}) становится равной
%$f_{\pi}^2m_{\pi}^2$~\cite{2a}, что ведет к логарифмической
%расходимости на верхнем пределе в~(\ref{c}). Однако, она может
%быть удалена путем переопределения кваркового массового
%члена~\cite{42} (см. также~\cite{de}).}
\begin{equation}
\label{pisum}
(m_{\pi^{+}}^{2}-m_{\pi^{0}}^{2})|_{em}=\frac{3}{4}\cdot
\frac{\alpha_{em}}{\pi
f_{\pi}^2}\cdot\sum_{k=1}^{\infty}\left\{f_{A,k}^2m_{A,k}^4 \ln
m_{A,k}^2-f_{V,k}^2m_{V,k}^4\ln m_{V,k}^2\right\}\,.
\end{equation}
Inserting in (\ref{pisum}) the two-resonance ansatz ($k=1$) for
VA-correlators, one comes to the following equation~\cite{37}:
\begin{equation}
\label{mpi} \Delta
m_{\pi}|_{em}^{(2)}\equiv(m_{\pi^{+}}-m_{\pi^{0}})|_{em}^{(2)}=
\frac{3\alpha_{em}}{4\pi(m_{\pi^{+}}+m_{\pi^{0}})}
\cdot\frac{m_{a_1}^{2}m_{\rho}^{2}}{m_{a_1}^{2}-m_{\rho}^{2}}
\ln\!\frac{m_{a_1}^{2}}{m_{\rho}^{2}}\,.
\end{equation}
Using the Weinberg relation $m_{a_1}=\sqrt{2}\,m_{\rho}$ one finds $\Delta
m_{\pi}|_{em}^{(2)}=5.21$~MeV. The experimental value is~\cite{31}:
$(m_{\pi^{+}}-m_{\pi^{0}})|_{exper}=4.59$~MeV.
In fact, the Weinberg relation is not exact.
Substitution in~(\ref{mpi}) of the physical mass for $a_1$-meson
$m_{a_1}=1230$~MeV gives estimation $\Delta
m_{\pi}|_{em}^{(2)}=5.79$~MeV (relative difference with the experimental value is 26$\%$).
%Анализ причин такого согласования можно найти,
%например, в~\cite{42, de}.
Notice that there are other contributions to $m_{\pi^{+}}-m_{\pi^{0}}$,
specifically, caused by the isospin symmetry breaking, i.e. an inequality of $u$ and $d$
quark masses~\cite{55, 56, 52} in the QCD. Their total magnitude is~\cite{32}: $\Delta
m_{\pi}|_{QCD}=0.17\pm0.03$~MeV. The correction of order $1/N_c$
does not exceed 7$\%$, as it follows from~\cite{mous, 39a}.
The effect of weak interactions is less than 1$\%$~\cite{elw}. Thus, the contribution
of electromagnetic interactions only turns out to be:
\begin{equation}
\label{mpiem}
\Delta m_{\pi}|_{em}=4.42\pm0.03\;\mbox{MeV}\,,
\end{equation}
and we will make comparison just with~(\ref{mpiem}) (as it is done, e.g. in~\cite{3a, 38}).
In such a way, the relative discrepancy of the result~(\ref{mpi}) with~(\ref{mpiem}) amounts
to 31$\%$ for the two-resonance ansatz. In the next section we will try to estimate this
difference taking into consideration higher meson resonances in the vector and axial-vector
channels.

\section{Calculation of electromagnetic $\pi$-meson mass difference with account of higher V,A-meson
resonances}

Now we proceed to the calculation of the electromagnetic $\pi$-meson mass difference
$\Delta m_{\pi}|_{em}$ in the case with two vector and two axial-vector resonances,
i.e. within the framework of the so called four-resonance ansatz. The utilization of the
Weinberg sum rules~(\ref{W1}),(\ref{W2}) made it possible to eliminate the
parameters $f_{\rho}$ and $f_{a_1}$ from~(\ref{mpi}). In the four-resonance case one has
three sum rules of chiral symmetry restoration (CSR) with the four unknown parameters
$f_{\rho},f_{a_1},f_{\rho'},f_{a_1'}$. The problem can be solved in a self-consistent way
by applying an approximate inequality $m_{a_1'}\gtrsim m_{\rho'}$, which follows
from properties of the mass spectrum obtained, in particular, in~\cite{20, we}. Let us
introduce a presumably small parameter
$\delta_{m}\equiv \frac{m_{a_1'}^{2}-m_{\rho'}^{2}}{m_{\rho'}^2}\ll1$.
%Таким
%образом, две переменные $f_{\rho'},f_{a_1'}$ эффективно
%превращаются в одну $\delta_{f}\equiv f_{\rho'}^{2}-f_{a_1'}^{2}$,
%в результате чего правила сумм ВКС становятся разрешимыми.
Saturating the correlators in~(\ref{c}) by four resonances and retaining only
the first power of~$\delta_{m}$ one obtains for the~$\Delta m_{\pi}|_{em}^{(4)}$:
$$
(m_{\pi^{+}}^{2}-m_{\pi^{0}}^{2})|_{em}^{(4)}\simeq\frac{3}{4}\cdot\frac
{\alpha_{em}}{\pi f_{\pi}^{2}}\times
$$
\begin{equation}
\label{mpi2}
\left\{f_{a_1}^{2}m_{a_1}^{4}\ln m_{a_1}^{2}-
f_{\rho}^{2}m_{\rho}^{4}\ln
m_{\rho}^{2}-\delta_{f}m_{\rho'}^{4}\ln m_{\rho'}^{2}+\epsilon
m_{\rho}^{2}(1+2\ln m_{\rho}^{2})\right\},
\end{equation}
where the unknown parameters~$f_{\rho}^{2},f_{a_1}^{2}$ and
$\delta_{f}\equiv f_{\rho'}^{2}-f_{a_1'}^{2}$ should be computed from the CSR sum rules
within a four-resonances consideration. Namely:
$$
m_{\rho}^{2}f_{\rho}^{2}-m_{a_1}^{2}f_{a_1}^{2}+m_{\rho'}^{2}\delta_{f}=
f_{\pi}^{2}+\epsilon
$$
\begin{equation}
\label{s}
m_{\rho}^{4}f_{\rho}^{2}-m_{a_1}^{4}f_{a_1}^{2}+m_{\rho'}^{4}\delta_{f}=
2m_{\rho'}^{2}\epsilon
\end{equation}
$$
m_{\rho}^{6}f_{\rho}^{2}-m_{a_1}^{6}f_{a_1}^{2}+m_{\rho'}^{6}\delta_{f}=
-4\!<\!\bar{q}q\!>^{2}+3m_{\rho'}^{4}\epsilon\,,
$$
where $\epsilon\equiv f_{a_1'}^{2}m_{\rho'}^2\delta_{m}$.

%Систему~(\ref{s}) можно решить аналитически, но ответы при этом
%довольно громоздкие.
Using experimental values for: $m_{\rho}=770$~MeV,
$m_{a_1}=1230\pm40$~MeV,\\ $m_{\rho'}=1465\pm25$~MeV, $f_{\pi}=93$~MeV
as well as bearing in mind an averaged value for the quark condensate
$<\!\bar{q}q\!>=-235\pm15$~$(\mbox{MeV})^{3}$ and a model estimation for a small parameter
$\epsilon$ (for example, from~\cite{we} and the condition $f_{a_1'}\lesssim f_{\rho'}$)
one can obtain from~(\ref{mpi2}) and~(\ref{s}):
\begin{equation}
\label{answ}
f_{\rho}\approx 0.18\;\qquad
f_{a_1}\approx 0.11\;\qquad
f_{\rho'}^{2}-f_{a_1'}^{2}\approx 0.0034\,,
\end{equation}
and for the electromagnetic $\pi$-meson mass difference $\Delta m_{\pi}|_{em}$:
\begin{equation}
\label{e}
\Delta m_{\pi}|_{em}^{(4)}\approx 3.85\pm0.16\;\mbox{MeV}.
\end{equation}
Taking into account the correction due to the quark condensate improves a result for $\Delta
m_{\pi}|_{em}^{(4)}$ by 5$\%$. The relative difference with~(\ref{mpiem}) amounts to 13$\%$
for the given ansatz. Notice that increasing the quark condensate value leads to enlargement
$\Delta m_{\pi}|_{em}^{(4)}$ (for example, if \\$<\!\bar{q}q\!>=-300$~$(\mbox{MeV})^{3}$
then $\Delta m_{\pi}|_{em}^{(4)}=4.42$ MeV), and employing the value for
$f_{\pi}$ that it would have in the chiral limit $f_{\pi}=87$ MeV~\cite{32}
changes the result less than 1$\%$.
%Изменение же масс V,A-резонансов при
%переходе к этому пределу весьма мало~\cite{mous}.

As it follows from the paper~\cite{36}: $f_{\rho}=0.20\pm 0.01$ (from the decay
$\rho^{0}\rightarrow e^{+}e^{-}$); $f_{a_1}=0.10\pm 0.02$ (from the decay
$a_{1}\rightarrow\pi\gamma$). The constant $f_{\rho'}$ is not yet determined from the
experiment because the electromagnetic decay of $\rho'$-meson is strongly suppressed
by hadronic decay channels. Nevertheless, Eq.~(\ref{answ}) provides a lower estimate
for~$f_{\rho'}\gtrsim0.06$. Our numerical estimates for~$f_{\rho},f_{a_1}$ coincide with those
of the work~\cite{23}.

We are able to compute also the constant $L_{10}$ of effective chiral Lagrangian~\cite{32},
which is defined by the mean electromagnetic $\pi$-meson radius $<\!r_{\pi}^2\!>$ and the
axial-vector pion formfactor $F_A$ for the decay~$\pi\rightarrow e\nu\gamma$ (see, e.g.~\cite{nar}).
Using the relation:
\begin{equation}
L_{10}=-\frac{1}{16}\frac{d}{dp^2}\left(p^2\left(\Pi^{V}(p^{2})-
\Pi^{A}(p^{2})\right)\right)_{p^2=0}\,,
\end{equation}
as well as~(\ref{kor}) and~(\ref{resid}), one easily obtains for $L_{10}$~\cite{20}:
\begin{equation}
\label{lten}
L_{10}=\frac{1}{4}\left(\sum_{n}f_{A,n}^{2}-\sum_{n}f_{V,n}^{2}\right)\,,
\end{equation}
that for the case $n=1,2$ leads to the estimate for the
constant~$L_{10}\approx-6.0\cdot 10^{-3}$, which is
consistent with that of~\cite{53} from hadronic $\tau$-decays:
$L_{10}=-(6.36\pm0.09|_{exp}\pm0.16|_{th})\cdot10^{-3}$.

Let us consider now how the above scheme of calculation of~$\Delta
m_{\pi}|_{em}$ works in the case of taking into account new resonances in the CSR sum rules.
By virtue of CSR at high energies one may expect that if the
inequality~$m_{a_1'}^{2}-m_{\rho'}^{2}\ll m_{\rho'}^{2}$ is fulfilled
then~$m_{a_1''}^{2}-m_{\rho''}^{2}\ll m_{\rho''}^{2}$ holds as well. Since
the mass~$m_{\rho''}$ is known, the set of equations~(\ref{s}) acquires only one new variable
$f_{\rho''}^{2}-f_{a_1''}^{2}$. As an additional condition one can put the constant~$f_{a_1}$
equal to its experimental value: $f_{a_1}\approx0.10$. Then a numerical solution gives:
$\Delta m_{\pi}|_{em}^{(6)}\approx 3.94$~MeV (relative difference with~(\ref{mpiem})
amounts to~11$\%$), $f_{\rho}\approx 0.18$,
$f_{\rho'}^{2}-f_{a_1'}^{2}\approx 0.0023$ (and, consequently,
$f_{\rho'}\gtrsim0.05$), $f_{\rho''}^{2}-f_{a_1''}^{2}\approx
0.0003$, $L_{10}\approx -6.2\cdot10^{-3}$. One can see that the addition of higher resonances
in calculation of the electromagnetic $\pi$-meson mass difference improves the results,
making them more consistent with the experimental value.

One can estimate the contribution of higher resonances to~(\ref{mpi}) in a
different way. Namely, taking into account in~(\ref{pisum}) the inequality:
\begin{equation}
\label{ooo}
\frac{m_{A,k}^2-m_{V,k}^2}{m_{V,k}^2}\ll1,\qquad k>1\,,
\end{equation}
which is a consequence of CSR at high energies, we come to:
$$
\sum_{k=2}^n\left(m_{A,k}^4f_{A,k}^2 \ln
m_{A,k}^2-m_{V,k}^4f_{V,k}^2\ln m_{V,k}^2\right)\simeq
$$
\begin{equation}
\label{mpi3}
(m_{\rho}^4f_{\rho}^2-
m_{a_1}^4f_{a_1}^2)\ln\bar{m}_{V,n}^2+\sum_{k=2}^n\left(\frac{m_{A,k}^2}{m_{V,k}^2}-1\right)
m_{A,k}^4f_{A,k}^2,
\end{equation}
where we have introduced an averaged mass $\bar{m}_{V,n}$:
\begin{equation}
\sum_{k=2}^n\left(m_{V,k}^4f_{V,k}^2-m_{A,k}^4f_{A,k}^2\right)\ln
m_{V,k}^2\equiv
\ln\bar{m}_{V,n}^2\cdot\sum_{k=2}^n\left(m_{V,k}^4f_{V,k}^2-m_{A,k}^4f_{A,k}^2\right).
\end{equation}

Under the assumptions made and the values of VA-meson spectral characteristic
admitted in this work, the second term in~(\ref{mpi3}) is by two-three order of magnitude
less then the first one (at least if~$n$ is not large). Therefore, we may neglect it
from now on. The expression~(\ref{pisum}) can be cast into the form:
$$
\Delta m_{\pi}|_{em}^{(n)}=\frac{3}{4}\cdot \frac{\alpha_{em}}{\pi
f_{\pi}^2(m_{\pi^{+}}+m_{\pi^{0}})}\times
$$
\begin{equation}
\label{pisum2}
\left\{(m_{a_1}^4f_{a_1}^2 \ln
m_{a_1}^2-m_{\rho}^4f_{\rho}^2\ln
m_{\rho}^2)-(m_{a_1}^4f_{a_1}^2-m_{\rho}^4f_{\rho}^2)
\ln\bar{m}_{V,n}^2\right\}\,.
\end{equation}
The second term in~(\ref{pisum2}) represents a correction to~(\ref{mpi})
(Eq.~(\ref{mpi}) is written in a form where the constants~$f_{\rho}$
and~$f_{a_1}$ are eliminated by means of the one-channel Weinberg sum rules). Were the
second one-channel sum rule~(\ref{W2}) exactly valid, the last term would vanish.

If we suppose quite a good convergence of the CSR sum rules (such that an account
of resonances with~$k>2$ therein would not lead to an essential change of spectral characteristics
of mesons with~$k\leq2$) then, in practice,
the quantity~$\bar{m}_{V,n}$ differs from~$m_{\rho'}$ slightly within our approximation.
In this way we can put~$\ln\bar{m}_{V,n}\simeq\ln m_{\rho'}$ and the generalized
Eq.~(\ref{mpi}) then has the form:
\begin{equation}
\label{pisum3}
\overline{\Delta m_{\pi}}|_{em}\simeq\frac{3}{4}\cdot
\frac{\alpha_{em}}{\pi
f_{\pi}^2(m_{\pi^{+}}+m_{\pi^{0}})}\cdot\left\{m_{\rho}^4f_{\rho}^2
\ln\frac{m_{\rho'}^2}{m_{\rho}^2}-m_{a_1}^4f_{a_1}^2\ln\frac{m_{\rho'}^2}{m_{a_1}^2}\right\}\,,
\end{equation}
where a bar means the averaged mass approximation for higher meson resonances.
Unlike the previous method, the expression~(\ref{pisum3}) does not contain the quark condensate
whose value varies considerably in literature.

As it follows from the derivation of formula~(\ref{pisum3}) we should substitute
there those values of~$f_{\rho}$ and~$f_{a_1}$ which they have in a given 2n-ansatz
(since the Weinberg sum rules should be fulfilled). For example, in the four-resonance
case they are~$f_{\rho}\approx0.18$,~$f_{a_1}\approx0.11$ and
the relation~(\ref{pisum3}) gives the
result~$\overline{\Delta m_{\pi}}|_{em}^{(4)}=3.64$~MeV almost coinciding with that
of~(\ref{e}) without taking account of the quark condensate. Compared with the one-channel
consideration the last value is better. In addition, the advantage of
formula~(\ref{pisum3}) over~(\ref{mpi}) is obvious in case of direct substitution
for~$f_{\rho}$,~$f_{a_1}$ by their experimental values and variation of the following
quantities within their experimental bounds:~$f_{\rho}=0.20\pm0.01$, $f_{a_1}=0.10\pm0.02$,
$m_{a_1}=1230\pm40$~MeV. Then the relation~(\ref{mpi}) (more strictly, Eq.~(\ref{pisum})
with~$k=1$) brings an absurd estimate:
$\Delta m_{\pi}|_{em}=102_{-120}^{+160}$~MeV, related with a poor fulfilment of the
one-channel Weinberg sum rules. At the same time a result of Eq.~(\ref{pisum3})
happens to be quite acceptable: $\overline{\Delta m_{\pi}}|_{em}=7.4\pm3.3$~MeV. In such a way,
the second term in~(\ref{pisum2}) is of the same order of magnitude as the
first one. Consequently, the account of
higher meson resonances is of importance. One may expect from Eq.~(\ref{pisum2}) that
when proceeding from the four-resonance ansatz to the six-resonance one and etc., a correction induced
grows
slowly (in a right direction) as effectively the account of higher meson resonances with $k>2$
leads to a slight increasing of the averaged mass~$\bar{m}_{V,n}$ and a corresponding
contribution does enter~(\ref{pisum2}) with a negative sign. As a result, the transition from the
two-resonance ansatz to the four-resonance one changes a value of electromagnetic
$\pi$-meson mass difference to a right direction and the account of
higher meson resonances ($k>2$) does it likewise, but results in an
insignificant correction only.

\section{Summary}

In this work we presented two ways of taking into account both vector ($\rho',\rho'',...$)
and axial-vector ($a_1',...$) higher meson resonances in the calculation of
the electromagnetic $\pi$-meson mass difference $\Delta m_{\pi}|_{em}$. The approaches are
based on the idea of chiral symmetry restoration at high energies as well as the Operator
Product Expansion for the correlators of vector and axial-vector quark densities. All calculations
were made  in the chiral limit, in the large-$N_c$ approximation and use the
asymptotic freedom of QCD.

In the first case the calculation of $\Delta m_{\pi}|_{em}$ is
carried out within the four-resonance approximation, where apart
from $\rho$ and $a_1$ - mesons their first excitations,
namely~$\rho'$ and~$a_1'$ - mesons, are also taken into
consideration. The conventional Weinberg sum rules are
supplemented by the the third sum rule~(\ref{W3}) ensuing from the
Operator Product Expansion and the assumption $m_{a_1'}\gtrsim
m_{\rho'}$. As a result, the following estimate for the
electromagnetic $\pi$-meson mass difference was obtained: $\Delta
m_{\pi}|_{em}^{(4)}\approx 3.85\pm0.16$~MeV, which improves its
theoretical prediction with regard to its experimental value
$\Delta m_{\pi}|_{em}^{exp}=4.42\pm0.03$~MeV (where the correction
due to this isospin symmetry breaking was taken into account) by
18$\%$. The estimations on electromagnetic decay constants
of~$\rho$ and~$a_1$ - mesons were also obtained:
$f_{\rho}\approx0.18$, $f_{a_1}\approx0.11$, which are in a good
agreement with~\cite{36, 23}, where they were derived in different
model approaches. The calculation of the constant~$L_{10}$ of
effective chiral Lagrangian~\cite{32} yields in this work:
$L_{10}\approx-6.0\cdot10^{-3}$, which is in a good agreement with
the experimental date following from hadronic
$\tau$-decays~\cite{53}:
$L_{10}=-(6.36\pm0.09|_{exp}\pm0.16|_{th})\cdot10^{-3}$. Moreover
it was shown that the account of next resonances improves the
result insignificantly, by order of several percents.

The second approach represents an extension of Eq.~(\ref{mpi})
for~$\Delta m_{\pi}|_{em}^{(2)}$ towards higher resonances. First of all,
it turns out that
when we use experimental values for masses and decay constants of~$\rho$ and~$a_1$ - mesons,
the generalized Eq.~(\ref{pisum3}) works better than~(\ref{pisum}) in the
one-channel~($k=1$) case. Secondly, one can see from the generalized Eq.~(\ref{pisum2})
that the account of~$\rho'$ and~$a_1'$ - meson reduces discrepancy
and that inclusion of higher resonances~($k>2$) leads to a certain improvement as well.

Finally we remark that in the recent work~\cite{pe} an attempt was made to estimate from the
phenomenology a number of quantities, including~$\Delta m_{\pi}|_{em}$, with the help of two-point
correlators being saturated by an infinite number of resonances of relevant mesons.
In this work the mass spectrum of higher excitations was parametrized by a trajectory of
Regge-Veneziano's type, unlike~\cite{we}, where mass spectrum of vector meson resonances was
calculated within the framework of Quasilocal Quark Models~\cite{11}
without any preliminary assumptions on a form and structure of the spectrum of higher meson
resonances. The value obtained for the electromagnetic $\pi$-meson mass
difference is~$\Delta m_{\pi}|_{em}^{(\infty)}=3.2$ MeV, i.e. the discrepancy
with~(\ref{mpiem}) makes up~28$\%$, which indicates an unsatisfactory approximation of spectral
characteristics of vector meson excitations in~\cite{pe}.
%Исследование же зависимости спектральных характеристик высших
%мезонных резонансов как от числа каналов
%$n$, так и от тензорной структуры вершин взаимодействия,
%представляет собой отдельную задачу и требует
%самостоятельного ее рассмотрения.
A somewhat different approach including meson excitations and calculation of their
spectral characteristics, based on the Nambu-Jona-Lasinio model with separable four-quark
interactions, can be found in~\cite{volk}.

We express our gratitude to A.A. Andrianov for fruitful
discussions and comments. This work is supported by Grant RFBR
01-02-17152, INTAS Call 2000 Grant (Project │ 587), Russian
Ministry of Education Grant E00-33-208, The Program "Universities
of Russia: Fundamental Investigations" (Grant 992612) and Grant
for Young Scientists of Sankt-Petersburg M01-2.4K-194.


\begin{thebibliography}{99}

\bibitem{37} T. Das, G.S. Guralnik, V.S. Mathur, F.E. Low, J.E.
Young, $Phys.$ $Rev.$ $Lett.$ {\bf 18} (1967) 759.
% Electromagnetic mass difference of pions, P. 483-492
%% CITATION = PRLTA, 18, 759;%%
\bibitem{57} ALEPH collaboration, R. Barate et al., $Eur.$ $Phys.$ $J.$
{\bf C4} (1998) 409.
% Measurement of the axial-vector $\tau$ spectral functions and determination
% of $\alpha_s(M\tau^2)$ from hadronic $\tau$ decays, P. 409-431
%% CITATION = EPHJA, C4, 409;%%
\bibitem{57d} OPAL collaboration, K. Ackerstaff et al., $Eur.$ $Phys.$ $J.$
{\bf C7} (1999) 571; hep-ex/9808019.
% Measurement of the strong coupling constant $\alpha_s$ and the
% vector and axial-vector spectral functions in hadronic tau decays, 571-593
%% CITATION = HEP-EX 9808019;%%
\bibitem{53} M. Davier, A. Hocker, L. Girlanda, J. Stern, $Phys.$ $Rev.$
{\bf D58} (1998) 096014; hep-ph/9802447.
% Finite energy chiral sum rules and spectral functions, P. 096014-096023
%% CITATION = HEP-PH 9802447;%%
\bibitem{54} R. Socolow, $Phys.$ $Rev.$ {\bf B137} (1965) 1221.
% Departures from eightfold way. 3. Pseudoscalar-meson
% electromagnetic masses, P. 1221-1265
\bibitem{min} T. Minamikawa et al., $Prog.$ $Theor.$ $Phys.$
$Suppl.$, Nos. {\bf 37}, {\bf 38} (1966) 56; $Prog.$ $Theor.$ $Phys.$,
{\bf 61} (1979) 548 (and references therein).
% Electromagnetic Mass Difference of Lowest Lying Hadrons and Behaviour
% of Quarks in Hadrons, P. 548-558
\bibitem{42} R.D. Peccei, J. Sola, $Nucl.$ $Phys.$ {\bf B281} (1987) 1.
% A phenomenological analysis of the Weinberg
% sum rules and of the $\pi^+-\pi^0$ mass difference, P. 1-17
%% CITATION = NUPHA, B281, 1;%%
\bibitem{36} G. Ecker, J. Gasser, A. Pich, E. de Rafael, $Nucl.$ $Phys.$ {\bf
B321} (1989) 311.
% The role of resonances in Chiral Perturbation Theory, P. 311-342
%% CITATION = NUPHA, B321, 311;%%
\bibitem{3a} W.A. Bardeen, J. Bijnens, J.-M. Gerard, $Phys.$ $Rev.$ $Lett.$
{\bf 62} (1989) 1343.
% Hadronic matrix elements and the $\pi^+-\pi^0$ mass difference, 12 pp.
%% CITATION = PRLTA, 62, 1343;%%
\bibitem{41} D. Espriu, E. de Rafael, J. Taron, $Nucl.$ $Phys.$ {\bf
B345} (1990) 22: $erratum$ $ibid.$ {\bf B355} (1991) 278. %278-279
% The QCD effective action at long distances, P. 22-56
%% CITATION = NUPHA, B345, 22;%%
\bibitem{38} J. Bijnens, E. de Rafael, $Phys.$ $Lett.$ {\bf B273} (1991) 483.
% The $\pi^+-\pi^0$ mass difference in the QCD effective approach, 483-492
%% CITATION = PHLTA, B273, 483;%%
\bibitem{43} J. Donoghue, B. Holstein, D. Wyler, $Phys.$ $Rev.$ {\bf D47} (1993) 2089.
% Electromagnetic self-energies of pseudoscalar mesons and Dashen's theorem, P. 2089-2097
%% CITATION = PHRVA, D47, 2089;%%
\bibitem{44} J. Bijnens, $Phys.$ $Lett.$ {\bf B306} (1993) 343; hep-ph/9302217.
% Violations of Dashen's theorem, P. 343-349
%% CITATION = HEP-PH 9302217;%%
\bibitem{33} J. Bijnens, E. de Rafael, H. Zheng, $Z.$ $Phys.$ {\bf
C62} (1994) 437; hep-ph/9306323.
% Low-energy behaviour of two-point functions of quark currents, P. 437-454
%% CITATION = HEP-PH 9306323;%%
\bibitem{mous} B. Moussallam, hep-ph/9804271.
% Reanalysis of the Das et al. sum rule and
% application to chiral $O(p^4)$ parameters
%% CITATION = HEP-PH 9804271;%%
\bibitem{Wein1} S. Weinberg, $Phys.$ $Rev.$ $Lett.$ {\bf 18} (1967) 507.
% Precise Relations between the Spectra of Vector
% and Axial-Vector Mesons, P. 507-509
%% CITATION = PRLTA, 18, 507;%%
\bibitem{31} D.E. Groom et al., $Eur.$ $Phys.$ $Jour.$ {\bf C15} (2000) 1
(particle data tables).
% Review of particle physics. Particle Data Group.
%% CITATION = EPHJA, C15, 1;%%
\bibitem{ani} V.V. Anisovich, hep-ph/0110326.
%% CITATION = HEP-PH 0110326;%%
\bibitem{a11} C.A. Baker et al., $Phys.$ $Lett.$ {\bf B449} (1999) 114.
% Evidence for $A\,J(PC)=1++$ meson at 1640 MeV, 114-121
%% CITATION = PHLTA, B449, 114;%%
\bibitem{a12} V. Dorofeev, VES Collaboration, hep-ex/9905002.
% New results from VES
%% CITATION = HEP-EX 9905002;%%
\bibitem{a13} O.A. Zaimidoroga, $Fiz.$ $Elem.$ $Chast.$ $Atom.$ $Yadra$
{\bf 30}, (1999) 5; $Phys.$ $Part.$ $Nucl.$ {\bf 30} (1999) 1.
% Radial excitations of light quark systems
%% CITATION = PRNUE, 30, 1;%%
\bibitem{25} A.A. Andrianov, V.A. Andrianov, $Zap.$ $Nauch.$
$Sem.$ $POMI$ {\bf 245} (1996) 5; hep-ph/9705364;
% Determination of scalar meson masses in QCD inspired quark models
%% CITATION = HEP-PH 9705364;%%
A.A. Andrianov, D. Espriu, R. Tarrach, $Nucl.$ $Phys.$ {\bf B533} (1998) 429;
hep-ph/9803232;
% The extended chiral quark model and QCD, 429-472
%% CITATION = HEP-PH 9803232;%%
A.A. Andrianov, V.A. Andrianov, $Proc.$ $of$ $the$ $Int.$ $Worksh$ $on$ $Hadron$
$Physics$, Coimbra 1999, ed. by A.H. Blin et al., N.Y., AIP, 2000,
p.382; hep-ph/9911383;
% Extended nonchiral quark models confronting QCD, P. 328-337
%% CITATION = HEP-PH 9911383;%%
A.A. Andrianov, V.A. Andrianov, R.Rodenberg, $JHEP$ {\bf 06}, 3 (1999);
hep-ph/9903287;
% Composite two Higgs models and chiral symmetry restoration
%% CITATION = HEP-PH 9903287;%%
\bibitem{18} M.A. Shifman, A.I. Vainstein, V.I. Zakharov, $Nucl.$ $Phys.$ {\bf
B147} (1979) 385,
% QCD and resonance physics --- theoretical foundations, P. 385-447
%% CITATION = NUPHA, B147, 385;%%
448.
% QCD and resonance physics --- applications, P. 448-518
%% CITATION = NUPHA, B147, 448;%%
\bibitem{32} J. Gasser, H. Leutwyler, $Nucl.$ $Phys.$ {\bf B250} (1985) 465.
%% CITATION = NUPHA, B250, 465;%%
%\bibitem{39} M. Knecht, R. Urech, $Nucl.$ $Phys.$ {\bf B519} (1998) 329.
\bibitem{39a} M. Knecht, R. Urech, preprint CPT-97/P.3524; $Nucl.$ $Phys.$
{\bf B519} (1998) 329; hep-ph/9709348.
% Virtual photons in Low Energy $\pi-\pi$ Scattering, P. 329-360
%% CITATION = HEP-PH 9709348;%%
\bibitem{16} G. t'Hooft, $Nucl.$ $Phys.$ {\bf B72} (1974) 461.
% A planar diagram theory for strong interactions, P. 461-473
%% CITATION = NUPHA, B72, 461;%%
\bibitem{17} E. Witten, $Nucl.$ $Phys.$ {\bf B160} (1979) 57.
% Barions in the 1/N expansion, P. 57-115
%% CITATION = NUPHA, B160, 57;%%
\bibitem{20} M. Knecht, E. de Rafael, $Phys.$ $Lett.$ {\bf B424} (1998) 335;
hep-ph/9712457.
% Patterns of spontaneous chiral symmetry breaking in the large
% $N_c$ limit of the QCD-like theories, 335-342
%% CITATION = HEP-PH 9712457;%%
\bibitem{19} L.J. Reinders, H. Rubinstein, S. Yazaki, $Phys.$ $Rept.$ {\bf 127} (1985) 1.
% Hadron properties from QCD sum-rules, P. 1-97
%% CITATION = PRPLC, 127, 1;%%
%\bibitem{2a} E. Floratos, S. Narison, E. de Rafael, $Nucl.$ $Phys.$ {\bf B155} (1979) 115.
% Spectral-function sum-rules in Quantum Chromodynamics. 1. Charged currents sector,
% P. 115-149
%% CITATION = NUPHA, B155, 115;%%
%\bibitem{de} V. De Alfaro, S. Fubini, G. Furlan, C. Rossetti, $Currents$
% $in$ $Hadron$ $Physics$ (North-Holland, Amsterdam, 1973).
\bibitem{55} D. Gross, S.B. Treiman, F. Wilczek, $Phys.$ $Rev.$ {\bf D19} (1979) 2188.
% Light-quark masses and isospin violation, P. 2188-2196
%% CITATION = PHRVA, D19, 2188;%%
\bibitem{56} J. Gasser, $Ann.$ $Phys.$ {\bf 136} (1981) 62.
%  Hadron masses and the sigma commutator in light of
% Chiral Perturbation Theory, P. 62-112
%% CITATION = APNYA, 136, 62;%%
\bibitem{52} J. Gasser, H. Leutwyler, $Ann.$ $Phys.$ {\bf 158} (1984) 142.
% Chiral Perturbation Theory to one loop, P. 142-210
%% CITATION = APNYA, 158, 142;%%
\bibitem{elw} M. Knecht, S. Peris, E. de Rafael, $Phys.$ $Lett.$ {\bf B443} (1998) 255;%v2
hep-ph/9809594.
% The electroweak $\pi^+-\pi^0$
% mass difference and weak matrix elements in the $1/N_c$ expansion, P. 255-263
%% CITATION = HEP-PH 9809594;%%
\bibitem{we} A.A. Andrianov, V.A. Andrianov, S.S. Afonin, hep-ph/0101245.
% Vector mesons in Quasilocal Quark Models
%% CITATION = HEP-PH 0101245;%%
\bibitem{nar} S. Narison, hep-ph/0012019.
% V-A hadronic tau decays: a QCD laboratory
%% CITATION = HEP-PH 0012019;%%
\bibitem{23} A.A. Andrianov, D. Espriu, $JHEP$ {\bf 10} (1999) 022; hep-ph/9906459;
% Vector mesons in the extended chiral quark model
%% CITATION = HEP-PH 9906459;%%
A.A. Andrianov, D. Espriu, R. Tarrach, $Nucl.$ $Phys.$ $Proc.$ $Suppl.$
{\bf 86} (2000) 275; hep-ph/9909366.
% The extended chiral quark model confronts QCD, 275-278
%% CITATION = HEP-PH 9909366;%%
%\bibitem{59} S. Gottlieb et al., $Nucl.$ $Phys.$ {\bf A498} (1989) 435.
\bibitem{pe} M. Golterman, S. Peris, $JHEP$ {\bf 0101} (2001) 028; hep-ph/0101098.
% Large $N_c$ QCD meets Regge theory: the example of spin one two
% point functions, 17 pp.
%% CITATION = HEP-PH 0101098;%%
\bibitem{11} A.A. Andrianov, V.A. Andrianov, $Int.$ $J.$ $Mod.$
$Phys.$ {\bf A8} (1993) 1981;
%  Structura of effective fermion models in symmetry-breaking phase, P. 1981-1992
%% CITATION = IMPAE, A8, 1981;%%
$Theor.$ $Math.$ $Phys.$ {\bf 94} (1993) 3; % $Theor.$ $Mat.$ $Fiz.$ {\bf 94} (1993) 6-18
% Effective fermion models with dynamical symmetry breaking, 3-10
%% CITATION = TMPHA, 94, 3;%%
$Proc.$ $School-Sem.$ "$Hadrons$ $and$ $nuclei$ $from$
$QCD$", $Tsuruga/Vladivostok/Sapporo$ $1993$, Singapore: WSPC,
1994, pp. 341-353; hep-ph/9309297;
% Effective Fermion Models in Symmetry-Breaking Phase and Quantum Chromodynamics
%% CITATION = HEP-PH 9309297;%%
$Nucl.$ $Phys.$ $Proc.$ $Suppl.$ {\bf 39BC} (1995) 257;
A.A. Andrianov, V.~A.~Andrianov, V.L. Yudichev, $Theor.$ $Math.$ $Phys.$
{\bf 108} (1996) 1069; % $Theor.$ $Mat.$ $Fiz.$ {\bf 108} (1996) 276-293
% Fermion models with quasilocal interaction in the vicinity of the
% polycritical point, 1069-1082
%% CITATION = IMPHA, 108, 1069;%%
A.~A.~Andrianov, V.A. Andrianov, D. Espriu, R. Tarrach, hep-ph/0009199.
%% CITATION = HEP-PH 0009199;%%
\bibitem{volk} M.K. Volkov, $Phys.$ $Atom.$ $Nucl.$ {\bf 60} (1997) 1920;
%$Yad.$ $Fiz.$ {\bf 60} Nom. 11 (1997) 1094;
hep-ph/9612456;
% The pseudoscalar and vector excited mesons in the $U(3)\times U(3)$ chiral model
%% CITATION = HEP-PH 9612456;%%
M.K. Volkov, V.L. Yudichev, $Phys.$ $Part.$ $Nucl.$
{\bf 31} (2000) 282.

\end{thebibliography}
\end{document}